\documentclass[slac_one]{revtex4}
\usepackage{graphicx}
\usepackage{fancyhdr}
\usepackage{epsfig}
\usepackage{epstopdf}

\begin{document}

\title{ \quad\\[0.5cm]   \boldmath Measurements of CKM angle $\phi_3$ at BELLE}

%

\author{S. Bahinipati\\
(Belle Collaboration)}
\affiliation{University of Cincinnati, Cincinnati, OH, 45221, USA}

\begin{abstract}

We report recent results on $\phi_3$ measurement at the Belle collaboration. The analyses reported here are based on a large data sample that contains 657 million $B\overline{B}$ pairs collected with the Belle detector at the KEKB asymmetric-energy $e^+ e^-$ collider at the $\Upsilon(4S)$ resonance. 
\end{abstract}

\maketitle

\thispagestyle{fancy}


\newcommand{\bdkp}{$B^{+}\to DK^{+}$}
\newcommand{\bdskp}{$B^{+}\to D^{*}K^{+}$}
\newcommand{\bdsk}{$B^{\pm}\to D^{*}K^{\pm}$}
\newcommand{\bdk}{$B^{\pm}\to DK^{\pm}$}
\newcommand{\bddsk}{$B^{\pm}\to D^{(*)}K^{\pm}$}
\newcommand{\dkpp}{$\overline{D}{}^0\to K^0_S\pi^+\pi^-$}

\section{Introduction}
In the Standard Model (SM), quark flavour mixing occurs via the Cabibbo-Kobayashi-Maskawa (CKM) matrix~\cite{KM}. $CP$ violation in the SM occurs due to the presence of a complex phase in the CKM matrix. Precision measurements of the parameters of CKM matrix are of utmost importance to constrain the SM and measure the amount of $CP$ violation. The CKM parameter $\phi_3$ ($\gamma$), defined as $\phi_3 = arg (-V_{ud}{V_{ub}}^{*}/V_{cd}{V_{cb}}^{*})$ is CKM angle measured with least precision. We report the recent measurements of $\phi_3$ by the Belle collaboration based on a large data sample that contains 657 million $B\overline{B}$ pairs in this report.

\section{Measurement of $CP$ violation parameters using $B^0 (\overline{B}{}^0) \to D^{*\mp}\pi^{\pm}$  decays}
\begin{figure}[!htb]
 \includegraphics[width=10.0cm,clip]{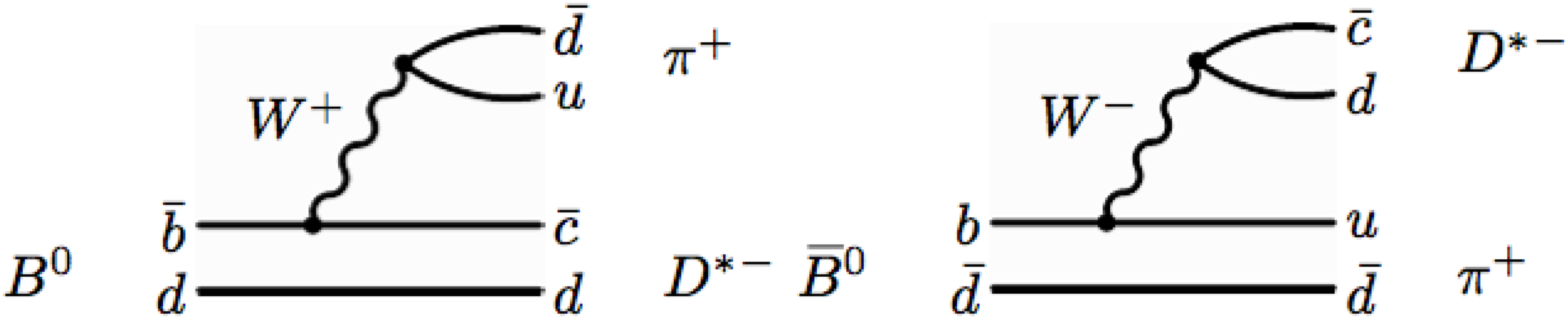}   
    \caption{
      Diagrams for 
       $B^0 \to D^{*-}\pi^+$ (left) and 
      $\overline{B}{}^0 \to D^{*-}\pi^+$ (right).
      Those for $\overline{B}{}^0 \to D^{*+}\pi^-$ and 
      $B^0 \to D^{*+}\pi^-$
      can be obtained by charge conjugation.}
      \label{fig:feynman}
\end{figure}

The study of the time-dependent decay rates of $B^0 (\overline{B}{}^0) \to D^{*\mp}\pi^{\pm}$ provides a theoretically clean method for extracting $\sin(2\phi_1+\phi_3)$~\cite{dunietz}, where $\phi_1$ and $\phi_3$ are angles of the CKM Unitarity Triangle. As shown in Fig.~\ref{fig:feynman}, this decay can be mediated by both Cabibbo-favoured (CFD) and doubly-Cabibbo-suppressed (DCSD) processes, whose  
amplitudes are proportional to ${V_{cb}}^{*}V_{ud}$ and ${V_{ub}}^{*}V_{cd}$ respectively, which have a relative weak phase $\phi_3$. 

\begin{figure}[!htb]
  \begin{center}
    \includegraphics[width=10.0cm,clip]{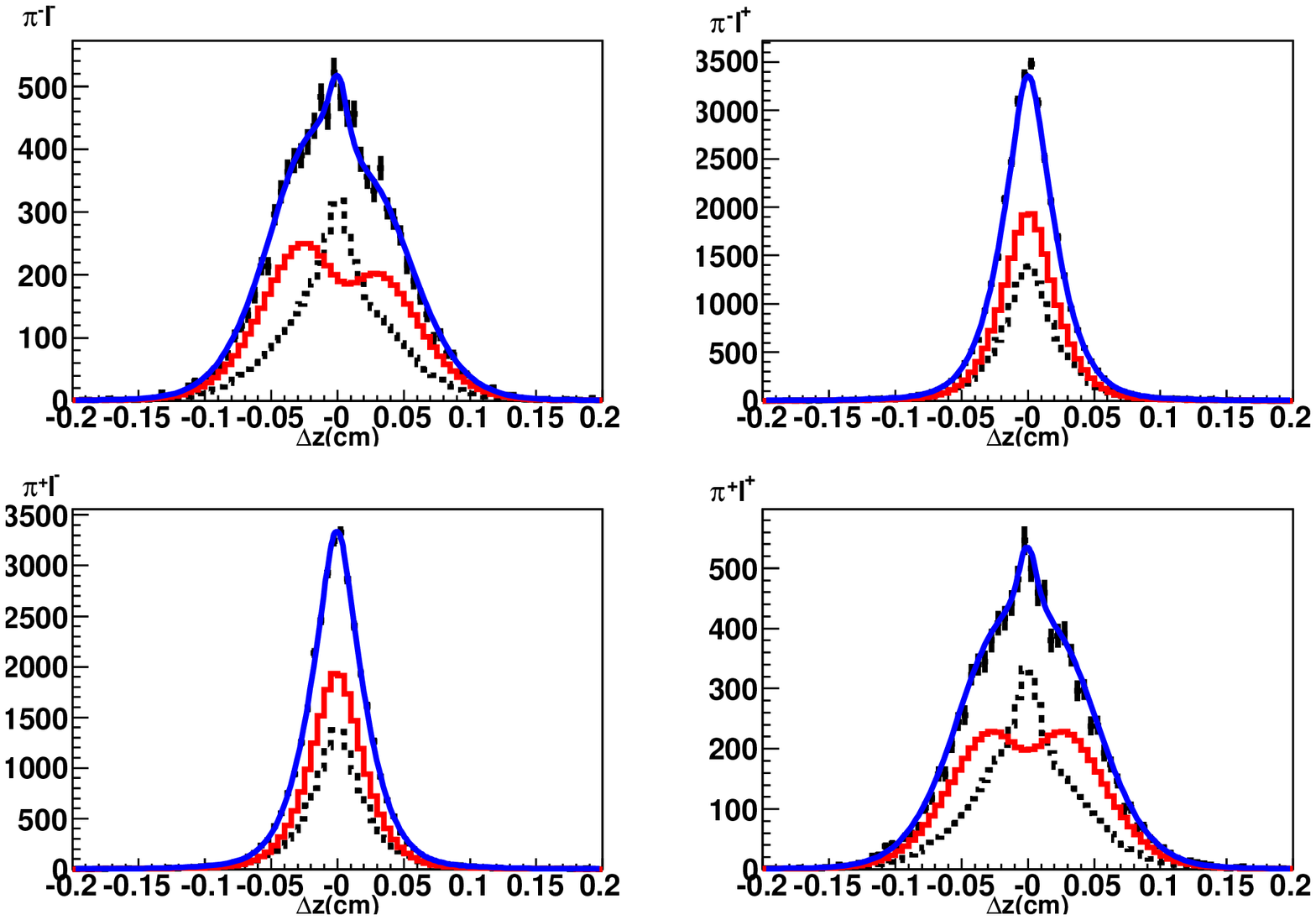}

  \end{center}
  \caption{
    \label{fig:myfit}
    $\Delta z$ distributions for 4 flavour-charge combinations: $\pi^{-}l^{-}$ (top left) , $\pi^{-}l^{+}$ (top right), $\pi^{+}l^{-}$ (bottom left), $\pi^{+}l^{+}$ (bottom right). The fit result is superimposed on the data (blue line).
    The signal and background components are shown as the red and dotted black curves, respectively.
   }
\end{figure}

The time-dependent decay rates are given by~\cite{fleischer}
\begin{eqnarray}
P(B^{0} &\to& D^{(*)\pm} \pi^\mp) = \frac{1}{8\tau_{B^0}} 
                    e^{-|\Delta t|/\tau_{B^0}} 
   \times \left[
      1 \mp C \cos (\Delta m \Delta t) - S^\pm \sin (\Delta m \Delta t) 
    \right],  \nonumber \\
P(\overline{B}{}^0 &\to& D^{(*)\pm} \pi^\mp) =  \frac{1}{8\tau_{B^0}}  
                  e^{-|\Delta t|/\tau_{B^0}} 
  \times    \left[
      1 \pm C \cos (\Delta m \Delta t) + S^\pm \sin (\Delta m \Delta t) 
    \right]. 
        \label{eq:evol}  
  \end{eqnarray}
Here $\Delta t$ is the difference between the time of the decay and the 
time that the flavour of the $B$ meson is tagged,  
$\tau_{B^0}$ is the average neutral $B$ meson lifetime, 
$\Delta m$ is the $B^0$-$\overline{B}{}^0$ mixing parameter, and 
$C = \left( 1 - R_{D^*\pi}^2 \right) / \left( 1 + R_{D^*\pi}^2 \right)$, 
where $R_{D^*\pi}$ is the ratio of the magnitudes between the DCSD and CFD 
(we assume the magnitudes of both the CFD and  DCSD amplitudes are the
same for $B^0$ and $\overline{B}{}^0$ decays). 
The $CP$ violation parameters are given by $S^{\pm} = -R_{D^*\pi} \sin(2\phi_1+\phi_3 \pm \delta_{D^*\pi})/
                \left( 1 + R_{D^*\pi}^2 \right)$ for $D^* \pi$, where $\delta$ is 
the strong phase difference between CFD and DCSD. Since the predicted value of $R_{D^*\pi}$ is small, $\sim$ 
$0.02$~\cite{csr}, we neglect terms of ${\cal O}\left( R_{D^*\pi}^2 \right)$ 
(and hence take $C = 1$).  The strong phase $\delta$ for $D^*\pi$ is predicted to be small~\cite{fleischer,wolfenstein}. Since $R_{D^*\pi}$ is expected to be suppressed, the amount of $CP$ violation in $D^*\pi$ decays, which is proportional to $R_{D^*\pi}$, is expected to be small and a large data sample is needed in order to
obtain sufficient sensitivity. We employ a partial 
reconstruction technique~\cite{zheng} for the $D^* \pi$ analysis, wherein the signal is distinguished from background on the basis of 
kinematics of the `fast' pion ($\pi_f$) from the decay $B \to D^* \pi$,
and the `slow' pion from the subsequent decay of  $D^* \to D \pi$;
the $D$ meson is not reconstructed at all. In order to tag the flavour of the associated $B$ meson,
we require the presence of a high-momentum lepton ($l$), required to have momenta in the range
$1.1 \ {\rm GeV}/c < p_{l} < 2.3 \ {\rm GeV}/c$ in the event. 
We perform a simultaneous unbinned fit to the 
same-flavour (SF) events, in which $\pi_f$ and $l$
have the same charge, and opposite-flavour (OF) events, in which the $\pi_f$ and $l$ have the opposite charge \cite{dstpi}. The results are shown in Fig.~\ref{fig:myfit}.
 We obtain the $CP$ violation parameters as $S^+  = +0.057\pm 0.019 \pm 0.012$, $S^-  =  +0.038 \pm 0.020 \pm 0.010$, where the first errors are statistical and the second errors are systematic.

\section{Measurement of $\phi_3$ from $B^\pm \to DK^\pm$ decays }
\begin{figure}[!htb]
 \includegraphics[width=11.0cm,clip]{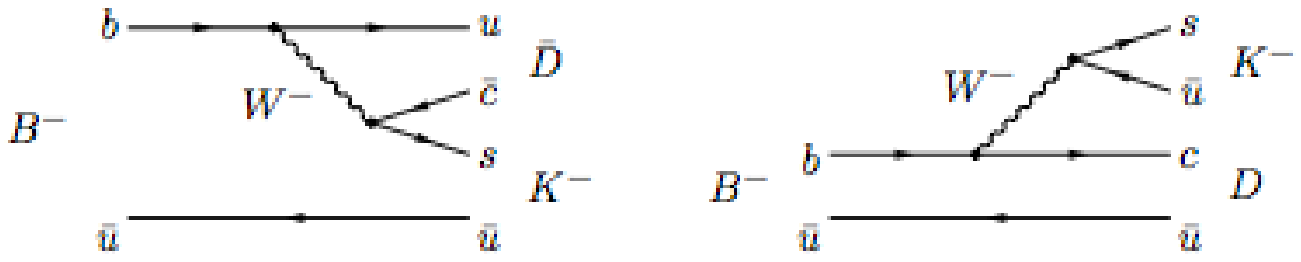}   
    \caption{
      Diagrams for 
       $B^- \to \overline{D} K^-$ and $B^- \to D K^-$}
      \label{fig:ads}
\end{figure}

 $CP$ asymmetries in the decays $B \to DK$ was first 
discussed by I. Bigi, A. Carter, and A. Sanda\cite{bigi}. Several methods have been proposed since then for $\phi_3$ measurement in such decays \cite{glw,dunietz,eilam,ads}. $\phi_3$ is accessible via interference of $V_{cb}$ and $V_{ub}$ amplitudes. The effects of $CP$ violation can be enhanced,  if the common final states 
of the $D^0$ and $\overline{D}{}^0$ decays following to $B^- \to D^0 K^-$ and $B^- \to \overline{D}{}^0 K^- (b \to c)$ are chosen so that the interfering amplitudes have comparable magnitudes (ADS method  \cite{ads}) (Fig.\ref{fig:ads}). The ratio of these interfering amplitudes, defined as $r_B$ is the ratio of the amplitudes of $B^- \to D^0 K^- (b \to u)$ and $B^- \to \overline{D}{}^0 K^- (b \to c)$ decays. The feasibility for measuring $\phi_3$ crucially depends on the size of $r_B$, which is predicted to be around 0.1-0.2 by taking a product of the ratio of the CKM matrix elements $|V_{ub} {V_{cs}}^{*} /V_{cb} {V_{us}}^{*}|$ and the color suppression factor. 

\begin{figure}[!htb]
  \epsfig{figure=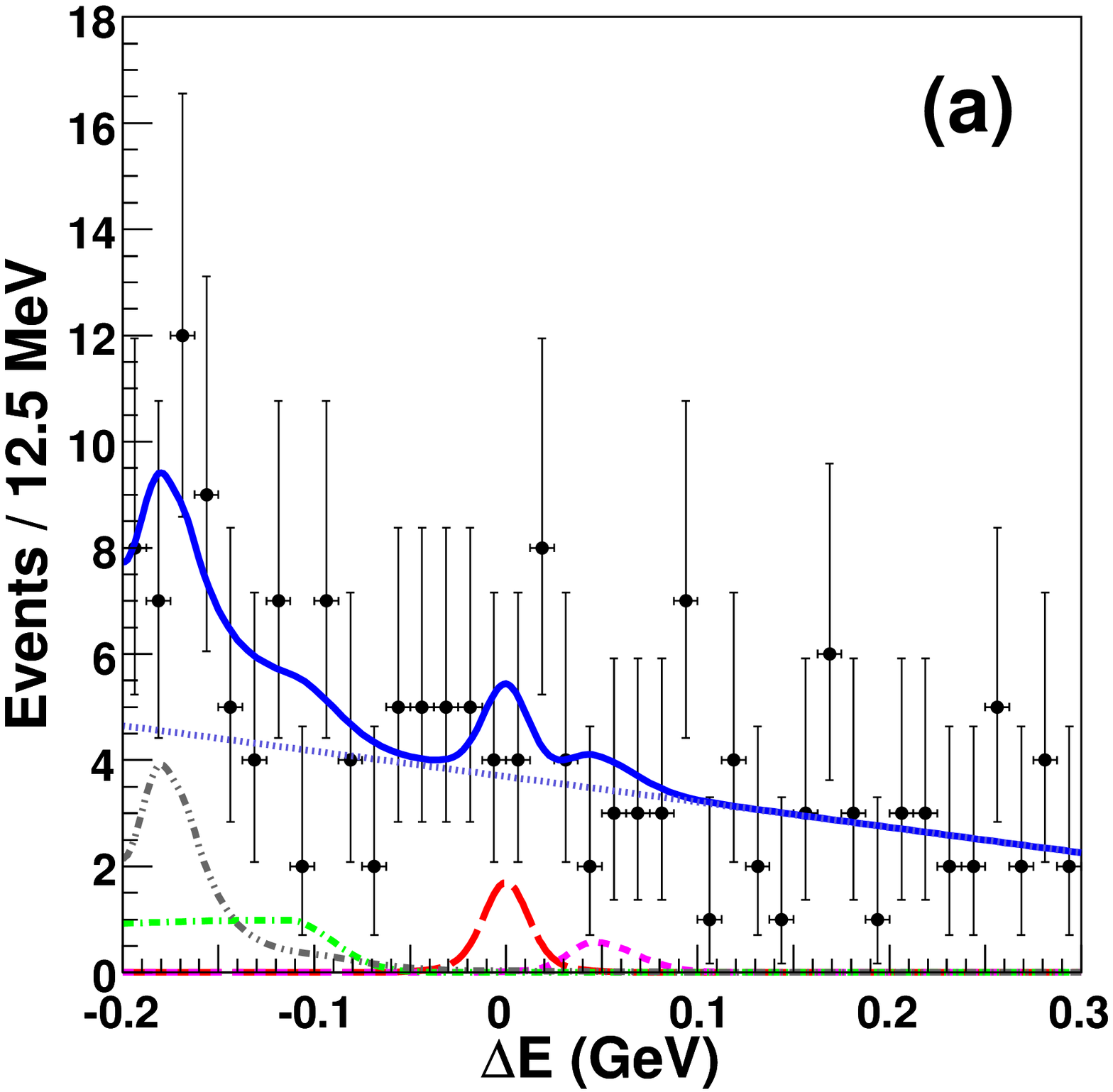,width=0.28\textwidth}
  \epsfig{figure=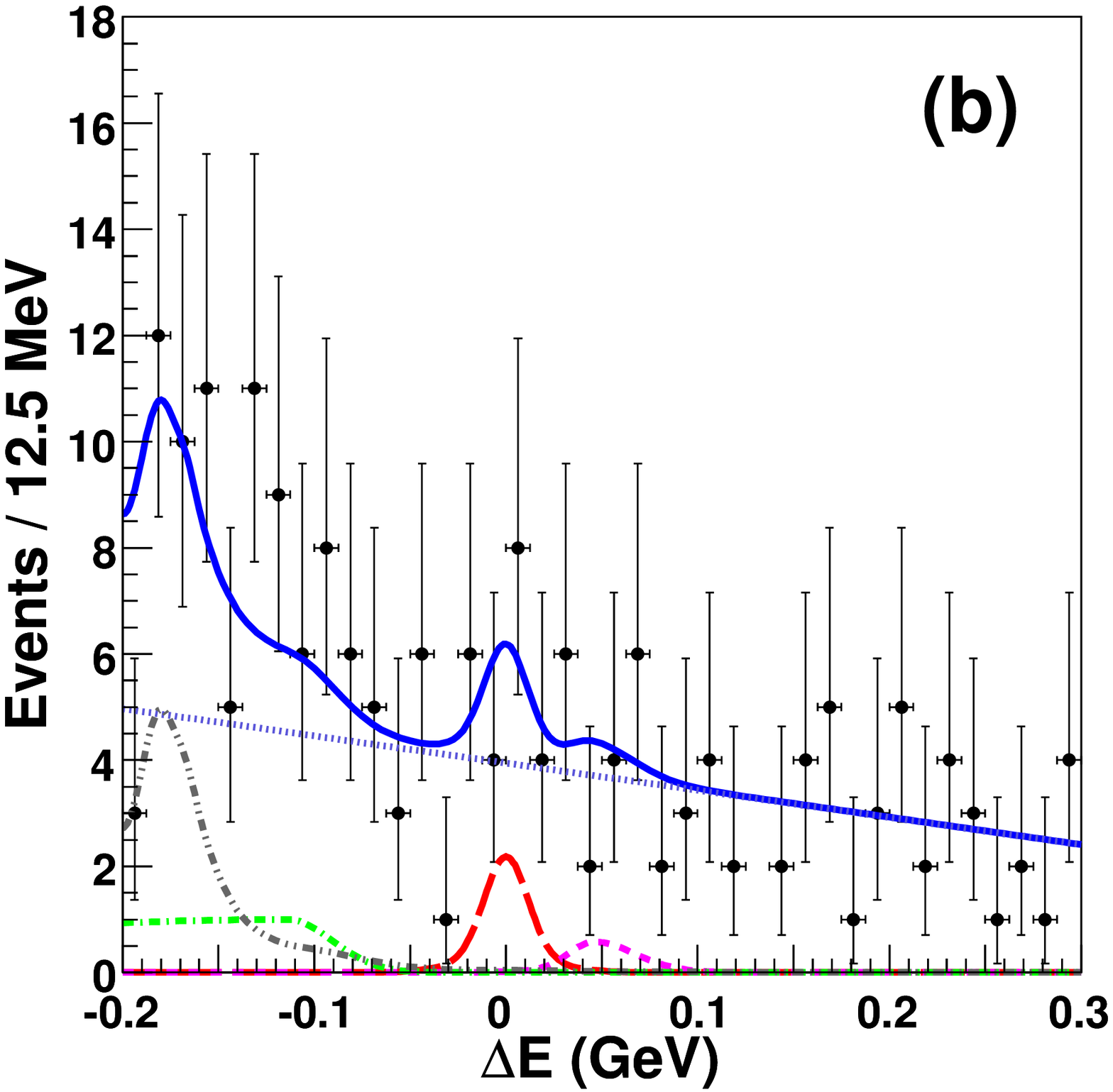,width=0.28\textwidth}

    \caption{The result of the fit to the energy difference between the signal candidate and the beam, $\Delta E$, for the mode (a) $B^- \to D K^-$, $D \to K^+ \pi^-$ and (b) $B^+ \to D K^+$, $D \to K^- \pi^+$. 
      }
      \label{fig:de}
\end{figure}

For the ADS method,  we define observables, the charge averaged rate ($R_{ADS}$) and the partial rate asymmetry ($A_{ADS}$) as $R_{ADS} = \frac {B(B^- \to [F ]_D K^- ) + B(B^+\to [\bar F ]_D K^+ )} {B(B^- \to [\bar F ]_D K^- ) + B(B^+ \to [F ]_D K^+)}$ and $A_{ADS} =  \frac {B(B^-\to [F ]_D K^- ) - B(B^+\to [\bar F ]_D K^+ )} {B(B^- \to [F ]_D K^- ) + B(B^+ \to [\bar F ]_D K^+)}$, where $[F ]_D$ indicates that the state $F$ originates from the $\overline{D}{}^0$ or $D^0$ meson. These observables are related to the physical parameters by $R_{ADS} = {r_B}^2 + {r_D}^2 + 2r_B r_D \cos (\delta_B + \delta_D ) \cos \phi_3 $ and $A_{ADS} = 2r_B r_D \sin (\delta_B + \delta_D ) \sin \phi_3 /R_{ADS}$ , where $r_D$ and $\delta_D$ are the ratio of the magnitudes and the strong phase difference of the $D$ decay amplitudes, respectively and $\delta_B$ is the ratio of the strong phase difference of the $B$ decay amplitudes. We obtain
 $R_{ADS} = [8.0^{+6.3}_{- 5.7}(stat) + 2.0 - 2.8 (syst)] \times 10^{-3}$ , 
$A_{ADS} = -0.13^{+ 0.97}_{- 0.88}(stat) \pm 0.26(syst)$ \cite{horii}. 
Although the signal (Fig.~\ref{fig:de}) is not significant, it allows to set an interesting upper limit on $r_B$. By taking a $+2\sigma$ variation on $r_D$ and conservatively assuming 
$\cos \phi_{3} \cos(\delta_B + \delta_D) = -1$, we obtain $r_B < 0.19$ at 90$\%$  confidence level.

\begin{figure}
  \epsfig{figure=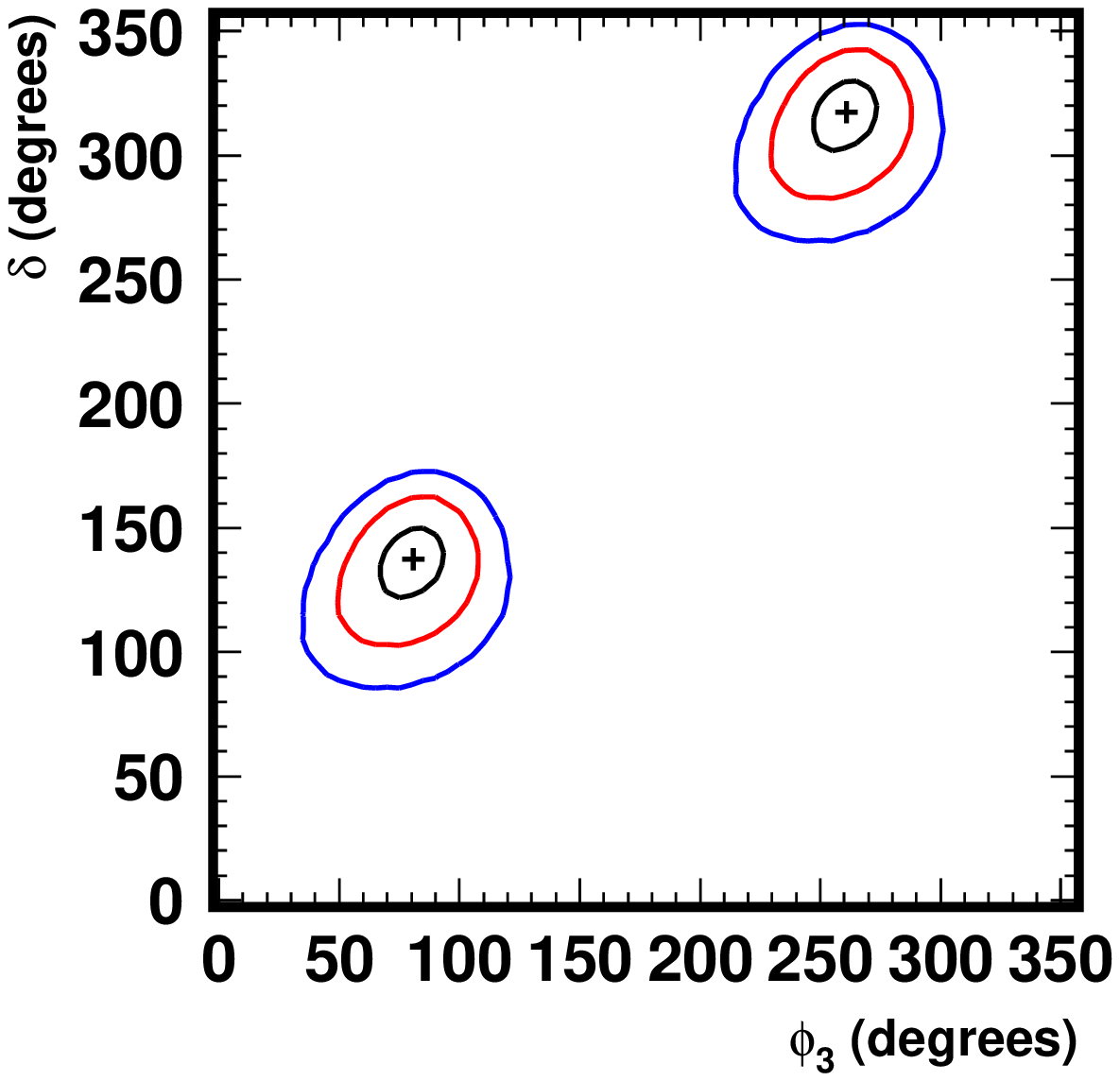,width=0.28\textwidth}
  \epsfig{figure=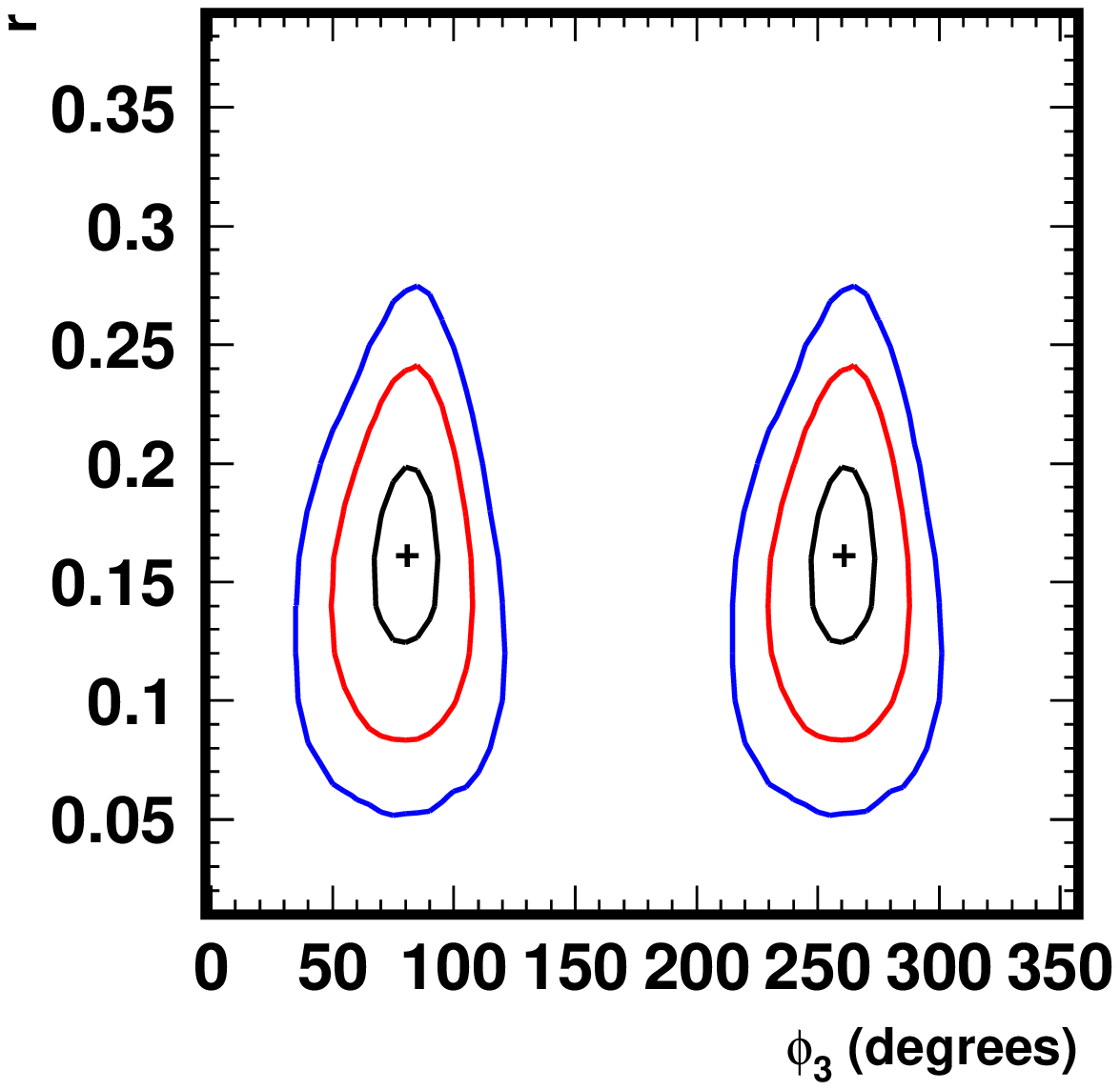,width=0.28\textwidth}\\
  \epsfig{figure=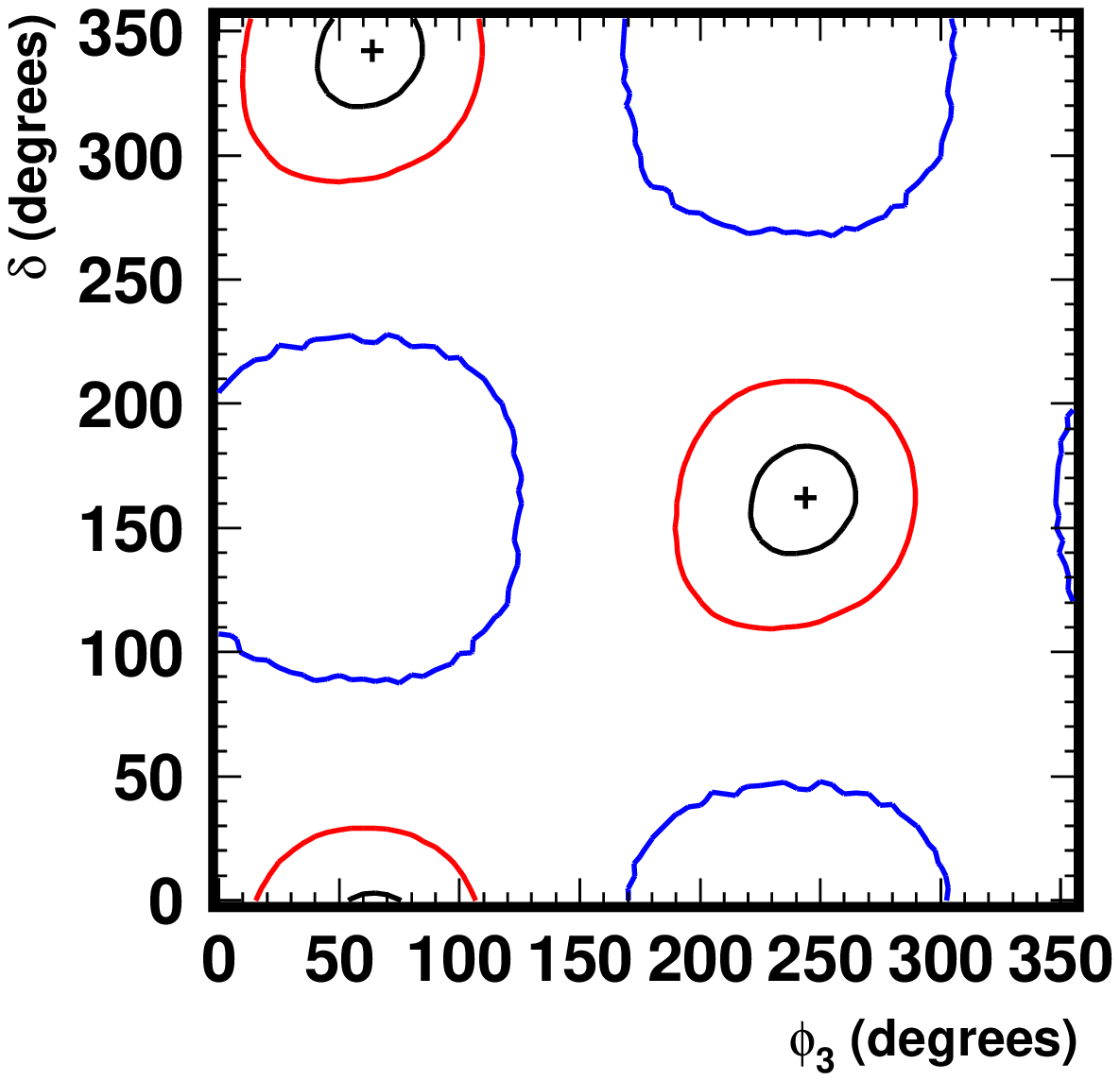,width=0.28\textwidth}
  \epsfig{figure=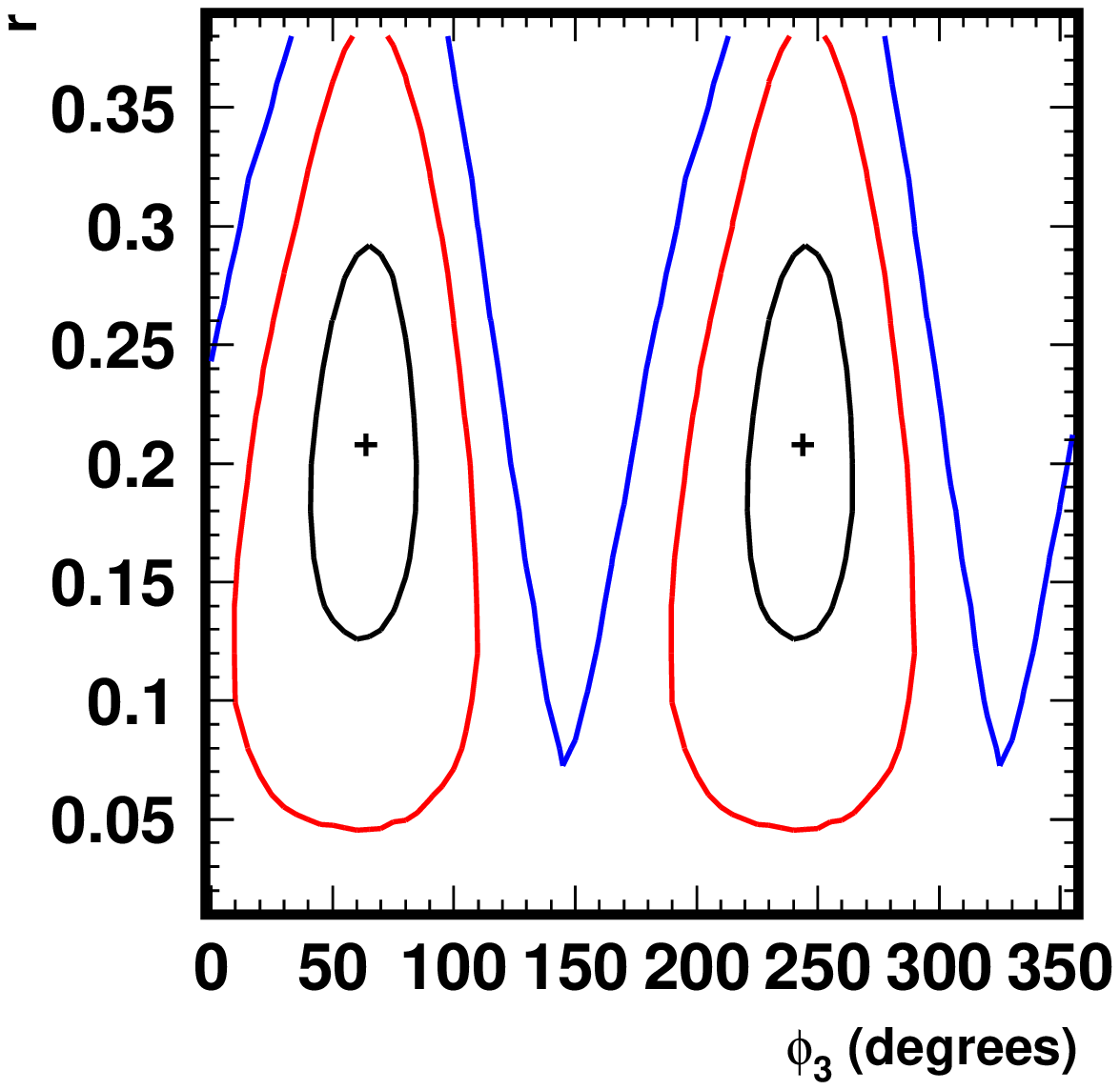,width=0.28\textwidth}
  \caption{Projections of confidence regions using statistical errors only for the \bdkp\ (top) and \bdskp\ (bottom) 
           mode onto the $(r_B, \phi_3)$ and $(\phi_3, \delta_B)$ planes. We show the 20\%, 74\% and 97\% confidence level regions, 
which correspond to one, two, and three standard deviations for a 
three-dimensional Gaussian distribution.
           }
  \label{cont_fc}
\end{figure}

Finally, the most effective constraint on $\phi_3$ comes from the $D \to K^0_s \pi^{+} \pi^{-}$ decay done using Dalitz analysis method\cite{giri,binp_dalitz}. We report results using two modes: \bdkp, and \bdskp\ with $D^{*}\to D\pi^0$, $D \to K^0_s \pi^{+} \pi^{-}$ as well as the corresponding charge-conjugate modes \cite{anton}. The weak parts of the amplitudes 
that contribute to the decay \bdkp\ 
are given by $V_{cb}^*V_{us\vphantom{b}}^{\vphantom{*}}\sim A\lambda^3$ 
(for the $\overline{D}{}^0 K^+$ final state) and
$V_{ub}^*V_{cs\vphantom{b}}^{\vphantom{*}}\sim A\lambda^3(\rho+i\eta)$ (for $D^0 K^+$). 
The two amplitudes interfere as the $D^0$ and $\overline{D}{}^0$ mesons decay
into the same final state $K^0_S \pi^+ \pi^-$. 
Assuming no $CP$ asymmetry in neutral $D$ decays, 
the amplitude of the neutral $D$ decay from \bdk\ 
as a function of Dalitz plot variables $m^2_+=m^2_{K^0_S\pi^+}$ and 
$m^2_-=m^2_{K^0_S\pi^-}$ is $M_{\pm}=f(m^2_{\pm}, m^2_{\mp})+re^{\pm i\phi_3+i\delta}f(m^2_{\mp}, m^2_{\pm})$, where $f(m^2_+, m^2_-)$ is the amplitude of the \dkpp\ decay,
$r_B$ is the ratio of the magnitudes of the two interfering amplitudes, 
and $\delta_B$ is the strong phase difference between them. The \dkpp\ decay amplitude $f$ can be determined
from a large sample of flavor-tagged \dkpp\ decays 
produced in continuum $e^+e^-$ annihilation. Once $f$ is known, 
a simultaneous fit of $B^+$ and $B^-$ data allows the 
contributions of $r_B$, $\phi_3$ and $\delta_B$ to be separated. The method\cite{belle_phi3_3} has a two-fold ambiguity: 
$(\phi_3,\delta_B)$ and $(\phi_3+180^{\circ}, \delta_B+180^{\circ})$
solutions cannot be separated. We always choose the solution 
with $0<\phi_3<180^{\circ}$. Figure~\ref{cont_fc} shows the projections of the three-dimensional 
confidence regions onto the $(r_B, \phi_3)$ and $(\phi_3, \delta_B)$ planes
for \bdk\ and \bdsk\ modes using statistical errors only. We perform a combined maximum likelihood fit to the two modes, and obtain\cite{anton}
$\phi_3=76^{\circ}\;^{+12^{\circ}}_{-13^{\circ}}
\mbox{(stat)}\pm 4^{\circ} \mbox{(syst)}\pm 9^{\circ}(\mbox{model})$. 
The statistical significance of $CP$ violation ($\phi_3\neq 0$) in 
our measurement is $(1-5.5\times 10^{-4})$, or 3.5 standard deviations. 

\section{Conclusion}
The precise measurement of $CKM$ angle $\phi_3$ is one of the most challenging, yet interesting pursuits in the $B$-factories. The Belle collaboration has performed $\phi_3$ extraction using several methods, and the most effective constraint on $\phi_3$ comes from the Dalitz plot analysis of the $K^0_S\pi^+\pi^-$ decay of the neutral $D$ meson produced in \bddsk\ decays. The recent Belle result for $B^- \to DK^-$ followed by $D \to K^+\pi^-$ brings a stringent upper limit on $r_B$, which is consistent with the result obtained by the Dalitz plot analysis. Results on time-dependent $CP$ asymmetries in $B \to D^{*\mp}\pi^{\pm}$ decays are also reported.

\end{document}